\documentclass[aps,prd,preprint,tightenlines,superscriptaddress,showpacs,byrevtex]{revtex4}

\usepackage{graphicx}
\usepackage{epsfig}
\usepackage{dcolumn}  
\graphicspath{{ps}}

\begin{document}

\vspace*{-3\baselineskip}
\resizebox{!}{3cm}{\includegraphics{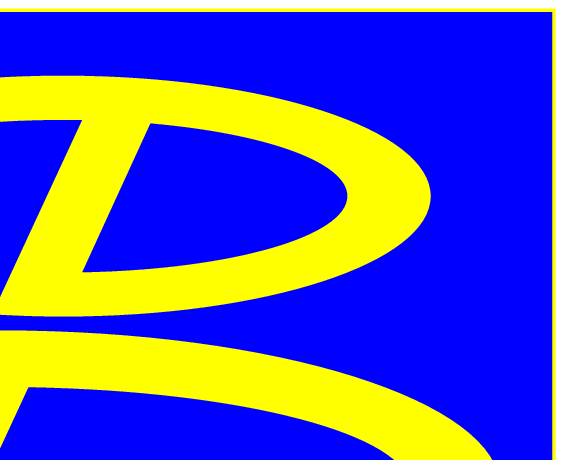}}

\preprint{\vbox{ 
                 \hbox{Belle preprint 2002-29 }
                 \hbox{KEK preprint 2002-84   }
                 \hbox{hep-ex/0208041}
}}

\title{ Study of Exclusive $B$ Decays to Charmed Baryons }

\date{\today}

\begin{abstract}

\noindent 

Using 29.1 fb$^{-1}$ of data accumulated at the $\Upsilon(4S)$ with
the Belle detector at KEKB, we have studied the decay modes
$\bar{B}^0\rightarrow\Lambda_{c}^{+}\bar{p}\pi^+\pi^-$,
$B^-\rightarrow{\Lambda_c^+}\bar{p}\pi^-$, and  
$\bar{B}^0\rightarrow{\Lambda_c^+}\bar{p}$.
We report branching
fractions of exclusive $B$ decays to charmed baryons with four-,
three- and two-body final states, including intermediate
$\Sigma_c^{++}$ and $\Sigma_c^{0}$ states.  We observed
$\bar{B}^0\rightarrow\Sigma_{c}(2455)^{++}\bar{p}\pi^-$ 
for the first time 
with a branching fraction of
$(2.38^{+0.63}_{-0.55}\pm0.41\pm0.62)\times 10^{-4}$ and observed
evidence for the two-body decay 
$B^-\rightarrow\Sigma_c(2455)^0\bar{p}$ with a branching fraction of
$(0.45^{+0.26}_{-0.19}\pm0.07\pm0.12)\times 10^{-4}$.  
We also set improved upper limits for the two-body decays $\bar{B}^0
\rightarrow\Lambda_c^+\bar{p}$ and $\bar{B}^-\rightarrow\Sigma_{c}(2520)^{0}\bar{p}$.

\end{abstract}

\affiliation{Budker Institute of Nuclear Physics, Novosibirsk}
\affiliation{Chiba University, Chiba}
\affiliation{Chuo University, Tokyo}
\affiliation{University of Cincinnati, Cincinnati OH}
\affiliation{University of Frankfurt, Frankfurt}
\affiliation{Gyeongsang National University, Chinju}
\affiliation{University of Hawaii, Honolulu HI}
\affiliation{High Energy Accelerator Research Organization (KEK), Tsukuba}
\affiliation{Hiroshima Institute of Technology, Hiroshima}
\affiliation{Institute of High Energy Physics, Chinese Academy of Sciences, Beijing}
\affiliation{Institute of High Energy Physics, Vienna}
\affiliation{Institute for Theoretical and Experimental Physics, Moscow}
\affiliation{J. Stefan Institute, Ljubljana}
\affiliation{Kanagawa University, Yokohama}
\affiliation{Korea University, Seoul}
\affiliation{Kyoto University, Kyoto}
\affiliation{Kyungpook National University, Taegu}
\affiliation{Institut de Physique des Hautes \'Energies, Universit\'e de Lausanne, Lausanne}
\affiliation{University of Ljubljana, Ljubljana}
\affiliation{University of Maribor, Maribor}
\affiliation{University of Melbourne, Victoria}
\affiliation{Nagoya University, Nagoya}
\affiliation{Nara Women's University, Nara}
\affiliation{National Lien-Ho Institute of Technology, Miao Li}
\affiliation{National Taiwan University, Taipei}
\affiliation{H. Niewodniczanski Institute of Nuclear Physics, Krakow}
\affiliation{Nihon Dental College, Niigata}
\affiliation{Niigata University, Niigata}
\affiliation{Osaka City University, Osaka}
\affiliation{Osaka University, Osaka}
\affiliation{Panjab University, Chandigarh}
\affiliation{Princeton University, Princeton NJ}
\affiliation{RIKEN BNL Research Center, Brookhaven NY}
\affiliation{Saga University, Saga}
\affiliation{University of Science and Technology of China, Hefei}
\affiliation{Seoul National University, Seoul}
\affiliation{Sungkyunkwan University, Suwon}
\affiliation{University of Sydney, Sydney NSW}
\affiliation{Tata Institute of Fundamental Research, Bombay}
\affiliation{Toho University, Funabashi}
\affiliation{Tohoku Gakuin University, Tagajo}
\affiliation{Tohoku University, Sendai}
\affiliation{University of Tokyo, Tokyo}
\affiliation{Tokyo Institute of Technology, Tokyo}
\affiliation{Tokyo Metropolitan University, Tokyo}
\affiliation{Tokyo University of Agriculture and Technology, Tokyo}
\affiliation{Toyama National College of Maritime Technology, Toyama}
\affiliation{University of Tsukuba, Tsukuba}
\affiliation{Utkal University, Bhubaneswer}
\affiliation{Virginia Polytechnic Institute and State University, Blacksburg VA}
\affiliation{Yonsei University, Seoul}
  \author{N.~Gabyshev}\affiliation{High Energy Accelerator Research Organization (KEK), Tsukuba} 
  \author{H.~Kichimi}\affiliation{High Energy Accelerator Research Organization (KEK), Tsukuba} 
  \author{K.~Abe}\affiliation{High Energy Accelerator Research Organization (KEK), Tsukuba} 
  \author{K.~Abe}\affiliation{Tohoku Gakuin University, Tagajo} 
  \author{T.~Abe}\affiliation{Tohoku University, Sendai} 
  \author{I.~Adachi}\affiliation{High Energy Accelerator Research Organization (KEK), Tsukuba} 
  \author{H.~Aihara}\affiliation{University of Tokyo, Tokyo} 
  \author{M.~Akatsu}\affiliation{Nagoya University, Nagoya} 
  \author{Y.~Asano}\affiliation{University of Tsukuba, Tsukuba} 
  \author{T.~Aso}\affiliation{Toyama National College of Maritime Technology, Toyama} 
% ...
\collaboration{The Belle Collaboration}
\maketitle \tighten

{\renewcommand{\thefootnote}{\fnsymbol{footnote}}}
\setcounter{footnote}{0}



Baryon production in flavored meson decays is unique 
to the $B$ meson system due to the heavy mass of the constituent b-quark.
Several studies of inclusive charmed baryon production in $B$ meson
decays~\cite{incl-lc} have been made and a large branching fraction
for $\bar{B}\rightarrow\Lambda_c^+ X$ of ($6.4\pm1.1$)\% has been
reported.  However, the mechanism is not well understood.  The
measured inclusive $\Lambda_{c}^{+}$ momentum spectra indicate that multi-body
final states are dominant in baryonic $B$ decays.  With a data sample
of 2.39 fb$^{-1}$, CLEO \cite{cleo-lamc} has studied exclusive charmed
baryonic decay modes and measured the branching fractions for
$\bar{B}^0\rightarrow\Lambda_{c}^+\bar{p}\pi^+\pi^-$ and
$B^-\rightarrow\Lambda_{c}^{+}\bar{p}\pi^-$.  They found no evidence
for $\bar{B}^0\rightarrow\Lambda_{c}^{+}\bar{p}$ and provided an
upper limit. So far, no observations of two-body decays 
have been reported.  On the other
hand, there are theoretical predictions for branching fractions of
two-body baryonic modes based on a pole model \cite{jarfi},
a QCD sum rule \cite{chernyak}, a diquark model \cite{ball}, and
a bag model \cite{cheng}.  The
predictions of the different models vary by an order of
magnitude, and experimental measurement can be used to
discriminate among them.
We have made a systematic study of exclusive charmed baryonic decays
of $\bar{B}^0$ and $B^-$ mesons into four-, three- and two-body final
states including $\Sigma_c^{++/0}$ intermediate resonances, 
by analyzing the 
$\Lambda_{c}^{+}\bar{p}\pi^+\pi^-$, ${\Lambda_c^+}\bar{p}\pi^-$
and ${\Lambda_c^+}\bar{p}$ final states.
Charge conjugate modes are included unless otherwise mentioned.
This analysis is based on a data sample
of 29.1 fb$^{-1}$ corresponding to $3.17\times{10^7}$ ${B\bar{B}}$ pairs.
The data were accumulated at the $\Upsilon(4S)$ resonance with the
Belle detector at the KEKB asymmetric collider of 3.5 GeV $e^+$ and
8.0 GeV $e^-$~\cite{kekb}.



The Belle detector is a large-solid-angle magnetic 
spectrometer that consists of a three-layer silicon vertex
detector (SVD), a 50-layer cylindrical drift chamber (CDC),
a mosaic of aerogel threshold \v{C}erenkov counters (ACC),
a barrel-like array of time-of-flight scintillation counters (TOF),
and an array of  CsI(Tl) crystals (ECL) located inside a 
superconducting solenoid coil that provides a 1.5~T magnetic field.
An iron flux return located outside the coil 
is instrumented to detect muons and $K_L$ mesons (KLM).  
The detector is described in detail elsewhere\ \cite{belle}.
We use a GEANT based Monte Carlo (MC) simulation to model the
response of the detector and determine the acceptance\ \cite{sim}.


In searches for the decay modes $\bar{B}^0\rightarrow \Lambda_{c}^+
\bar{p}~\pi^+ \pi^-$, 
$B^-\rightarrow\Lambda_{c}^{+}\bar{p}~\pi^-$, 
and $\bar{B}^0\rightarrow \Lambda_{c}^+ \bar{p}$,
the ${\Lambda_c^+}\rightarrow{pK^-\pi^+}$ decay mode is used.
Particle identification information from the CDC $dE/dx$, ACC and TOF
is used to provide a mass assignment
for each track. A likelihood ratio $LR(A,B)=L_A/(L_A+L_B)>0.6$ is
required to identify a particle as type $A$, where $B$ is the other
possible assignment among $\pi^{\pm}$, $K^{\pm}$ and $p(\bar{p})$.
Electron and muon candidate tracks are removed if their probabilities 
from the ECL, CDC $dE/dx$ and KLM are greater than 95\%.  Candidate
${\Lambda_c^+}$'s are tagged if the invariant mass of the $p$, $K^-$
and $\pi^+$ track combination is within 0.010~GeV/$c^2$ 
of the ${\Lambda_c^+}$ mass; 
tagged events are then examined for the three search
modes by adding $\bar{p}$, $\pi^-$, and $\pi^+$ tracks.
The width $\sigma_{\Lambda_c^+}$ is found to be 4.9~MeV/$c^2$,
consistent with the MC.

In order to select $\bar{B}$ meson candidates, 
we use the beam energy-constrained
mass and energy difference, which are defined as
$M_{\rm{bc}}=\sqrt{E_{\rm beam}^2-{(\sum\vec{p_i})}^2}$ and
$\Delta{E}=\sum{E_i}-E_{\rm beam}$ 
in the center-of-mass (CM) frame of the $e^+ e^-$ collision.  
$E_{\rm beam}$ is the beam energy, and $E_i$ and $\vec{p}_i$ are the energy and
momentum vector for the $i$-th daughter particle of a $B$ candidate.
$B$ candidates are selected with a loose cut to retain sideband
events by requiring $M_{\text{bc}}>5.2\,\mathrm{GeV}/c^2$ and
$|\Delta{E}|<0.2$~GeV.
A vertex-constrained fit for the three daughter tracks is carried out
at the ${\Lambda_c^+}$ vertex. For each decay mode, the virtual
${\Lambda_c^+}$ track and additional tracks are required to form a
good vertex.
If there are multiple candidates for both $\Lambda_c^+$ and $B$, the
candidate with the minimum
$\chi^2=\chi^2_{\Lambda_c^+}+\chi^2_{B}+(M_{\rm{bc}}-5.279)^2/\sigma_{M_{\rm{bc}}}^2$
is selected.  Here, $\chi^2_{\Lambda_c^+}$ and $\chi^2_{B}$ are the
$\chi^2$'s from the fits for the $\Lambda_c^+$ and $B$ vertices,
respectively, and 
$\sigma_{M_{\text{bc}}}$ is the MC value of the 
$M_{\text{bc}}$ width (2.8 MeV/$c^2$).  
Loose cuts on $\chi^2_{\Lambda_c^+}$ and $\chi^2_{B}$
are applied to remove background from tracks arising from
$K_S^0$ and $\Lambda$ decays.

Event selection requirements are optimized using signal MC events and
continuum background MC events consisting of $u\bar{u}$, $d\bar{d}$,
$s\bar{s}$, and $c\bar{c}$ quark-antiquark pairs generated with the
expected fractions.  To suppress the continuum background, we use a
Fisher discriminant constructed from 10 variables: 8 modified Fox Wolfram
moments \cite{sfw}, cos$\Theta_{B}$, and
cos$\Theta_{\Lambda_c^+}$.  Here, cos$\Theta_{B}$ is the cosine of
the direction of the $B$ meson with respect to the electron beam
direction, and cos$\Theta_{\Lambda_c^+}$ is the cosine of the
direction of the daughter $\Lambda_c^+$ with respect to the thrust
axis of the tracks not associated with the $B$ candidates.  Both quantities
are defined in the CM system.  A set of 10 coefficients for each mode
is optimized to maximize separation of the signal from the continuum
background.  The probability density functions for the signals 
and for the continuum, $P_{\rm sig}$ and
$P_{\rm con}$ respectively, are parameterized
with Gaussian functions for the three search modes and for the continuum
events. A cut on the likelihood ratio $R_{\rm sfw}=P_{\rm
sig}/(P_{\rm sig}+P_{\rm con})>0.6$ is applied to all decay modes.
In the MC simulation this cut removed 76\% of the continuum background 
while retaining 86\% of the signal for $\Lambda_c^+\bar{p}\pi^+\pi^-$.

\begin{figure*}[!htb]
\centering 
\begin{tabular}{lr}
\mbox{\psfig{figure=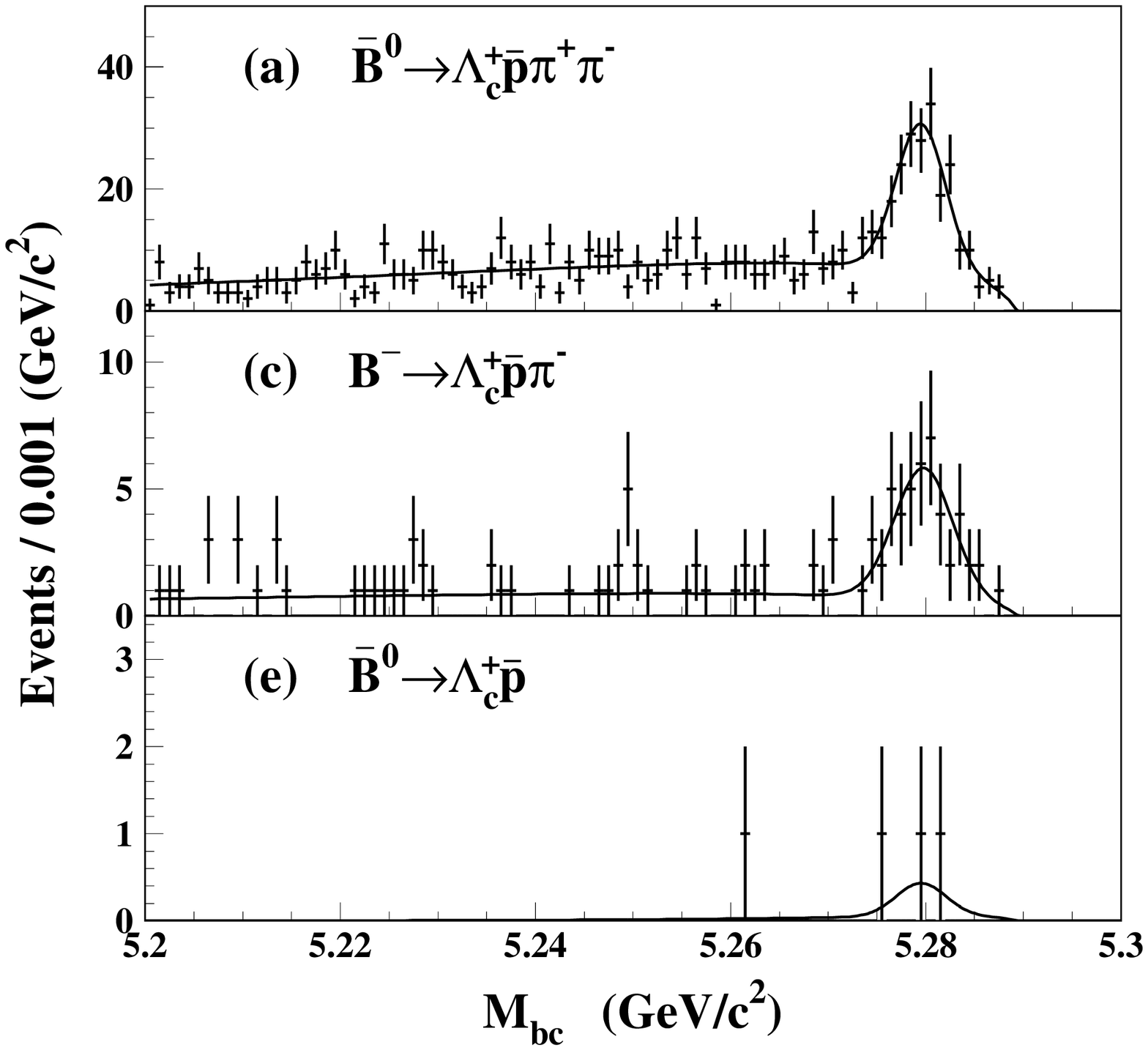,width=0.48\textwidth}}
& \mbox{\psfig{figure=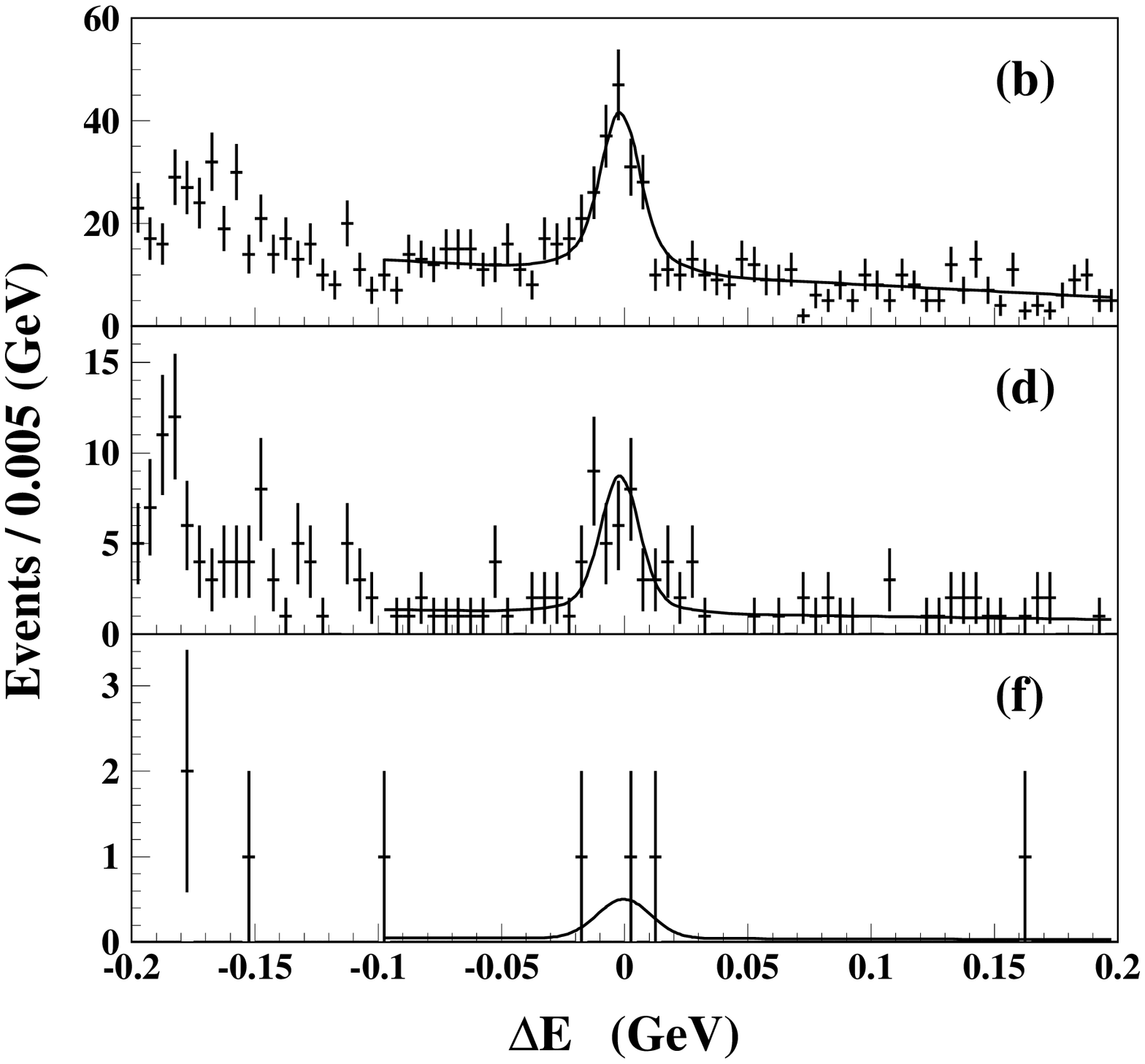,width=0.48\textwidth}} \\
\end{tabular}
\vspace{-0.30cm}
\caption{ $M_{\rm{bc}}$ distributions for $|\Delta{E}|<0.030$~GeV and
$\Delta{E}$ distributions for $M_{\rm{bc}}>5.270$~GeV/$c^2$: (a) and
(b) for $\bar{B}^0\rightarrow\Lambda_{c}^+\bar{p}\pi^+\pi^-$, (c) and
(d) for ${B^-}\rightarrow\Lambda_{c}^+\bar{p}\pi^-$, and
(e) and (f) for $\bar{B}^0\rightarrow\Lambda_{c}^+\bar{p}$.
Points with errors indicate the data and the curves indicate fits 
(see text for details).
}
\label{fig:fig1}
\end{figure*}

Figure~\ref{fig:fig1} shows the $M_{\rm{bc}}$ and $\Delta{E}$ distributions
for the three decay modes, after a tight cut is made in the 
($\Delta{E}$,$M_{\rm{bc}}$) variable not plotted. 
The $M_{\rm bc}$ background 
distributions are parameterized by the ARGUS function
\cite{ArgusF}, while a Gaussian is used for the signal. The $\Delta{E}$
distributions are fitted with a second-order polynomial for the background
and a double Gaussian for the signal. Here, the width parameters are fixed to
the values fitted to the signal MC events.
The mean and width of
$M_{\rm{bc}}$ in the data are found to be consistent with the MC
values of $5.279$~GeV/$c^2$ and $2.8$~MeV/$c^2$, respectively.
The width of $\Delta{E}$ is also consistent 
with the MC value ($9.9$~MeV) when fit to a single Gaussian. 
We obtain signal yields of
$154^{+17}_{-16}$ and $38.8^{+7.6}_{-7.0}$ from the fits to the
$M_{\rm{bc}}$ distributions (a) and (c), and $141^{+16}_{-15}$
and $30.2^{+7.0}_{-6.4}$ from the fits to the $\Delta{E}$ distributions
(b) and (d), respectively.  Here, we choose the asymmetric range of
$-0.100<\Delta{E}<0.200$~GeV to exclude feed-down from 
higher multiplicity modes with extra
pions; these produce the structure observed in the region
$\Delta{E}<-0.150$~GeV.  Since $M_{\rm{bc}}$ is used 
in the $\chi^2$ calculation for the best
candidate selection as described previously, 
we use the yields resulting from the fits to the
$\Delta{E}$ distributions to calculate branching fractions.

We observe $\bar{B}^0\rightarrow\Lambda_c^+\bar{p}\pi^+\pi^-$ and
$B^-\rightarrow\Lambda_c^+\bar{p}\pi^-$ signals.
For $\bar{B}^0\rightarrow\Lambda_c^+\bar{p}$ we find
a statistical significance of only $1.9\,\sigma$ from 
a fit to a Gaussian function for the signal with mean and width fixed 
to those from the signal MC, and a linear background function.
We thus set an upper limit of 6.1 events at the 90\% confidence 
level based on the likelihood function, 
using the Bayesian method with a prior uniform in the branching fraction.

Table~\ref{table:yield} summarizes the observed yields and branching fractions. 
Here, the detection efficiencies are calculated assuming nonresonant decays and
do not include the branching fraction
${\cal{B}}(\Lambda_c^+\rightarrow{pK^-\pi^+})$=($5.0\pm1.3$)\%~\cite{pdg2000}.
We assume the fractions of charged and neutral $B$ mesons to be equal in
the branching fraction calculations.
We include a correlated systematic error of 2\% per track for
tracking and particle identification.
Systematics due to the $\chi^2_{\Lambda_c^+}$,
$\chi^2_{B}$ and $R_{\rm sfw}$ cuts
are estimated by varying cut values.
The signal shape systematic error is evaluated 
from the variation in fit results obtained
with different-order polynomials used for 
the background and single and double
Gaussians used for the signal. The resulting total systematic errors for
$\Lambda_c^+\bar{p}\pi^+\pi^-$, $\Lambda_c^+\bar{p}\pi^-$ and
$\Lambda_c^+\bar{p}$ are 17.2\%, 14.8\% and 13.3\%, respectively.
Table~\ref{table:yield} shows the CLEO measurements 
renormalized to the same 
${\cal{B}}(\Lambda_c^+\rightarrow{pK^-\pi^+})$ for comparison.
Our branching fraction for 
$\bar{B}^0\rightarrow\Lambda_c^+\bar{p}\pi^+\pi^-$ is 
consistent with their measurement; however, our result for 
$B^-\rightarrow\Lambda_c^+\bar{p}\pi^-$ is somewhat lower (1.5$\,\sigma$).
We also set a more restrictive upper limit 
on $\bar{B}^0\rightarrow\Lambda_c^+\bar{p}$.

\begin{table*}[!htb]
\caption{
Branching fractions for
$\bar{B}^0\rightarrow\Lambda_c^+\bar{p}\pi^+\pi^-$,
$B^-\rightarrow\Lambda_c^+\bar{p}\pi^-$, and
$\bar{B}^0\rightarrow\Lambda_c^+\bar{p}$.
The errors are statistical, systematic, and a common error due to
the uncertainty in the value of 
${\cal{B}}(\Lambda_c^+\rightarrow{p K^- \pi^+})$.
The CLEO results are renormalized to
${\cal{B}}(\Lambda_c^+\rightarrow{pK^-\pi^+})=(5.0\pm1.3)\%$ \cite{pdg2000}
for comparison. 
}

%
\begin{ruledtabular}
\begin{tabular}{lccccc}
Mode & Efficiency (\%) & Yield & Significance & ${\cal{B}}~(\times10^{-4}$) & CLEO~($\times10^{-4}$) \\ \hline
$\bar{B}^0\rightarrow\Lambda_c^+\bar{p}\pi^+\pi^-$
& 8.07 & $141^{+16}_{-15}$ & 12.2 & $11.0^{+1.2}_{-1.2}\pm1.9\pm2.9$
& $11.7^{+4.0}_{-3.7}\pm2.7\pm3.0$ \\
$B^-\rightarrow\Lambda_c^+\bar{p}\pi^-$
&10.2 & $30.2^{+7.0}_{-6.4}$ & 6.0 & $1.87^{+0.43}_{-0.40}\pm0.28\pm0.49$ &
$5.5^{+2.0}_{-1.8}\pm1.0\pm1.4$ \\
$\bar{B^0}\rightarrow\Lambda_c^+\bar{p}$
& 12.9 & $2.4^{+2.1}_{-1.5}$ & 1.9 & $0.12^{+0.10}_{-0.07}\pm0.02\pm0.03$ &  \\
&  & \hspace{-5mm}$<~$6.1~(90\%~CL)\hspace{-5mm} & &  $<~$0.31~(90\%~CL) &  $<$~1.85~(90\%~CL) \\
\end{tabular}
\end{ruledtabular}
\label{table:yield}
\end{table*}

\begin{figure}[!htb]
\centering
\mbox{\psfig{figure=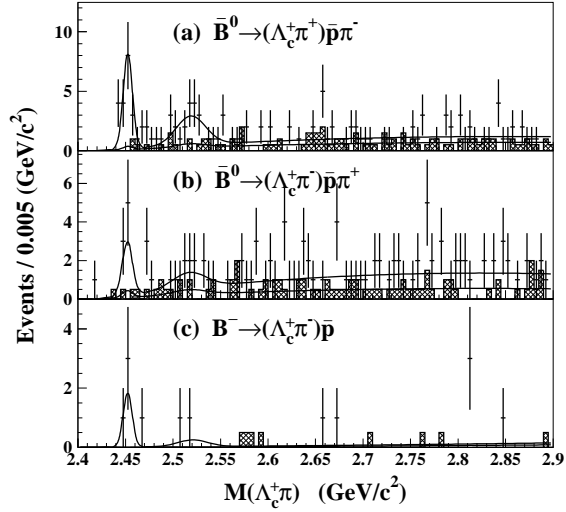,width=0.48\textwidth}}
\vspace{-0.30cm}
\caption{ Invariant mass distributions (a) $M(\Lambda_c^+\pi^+)$ and (b)
$M(\Lambda_c^+\pi^-)$ 
for $\bar{B}^0\rightarrow \Lambda_{c}^+ \bar{p} \pi^+ \pi^-$,
and (c) $M(\Lambda_c^+\pi^-)$ 
for ${B^-}\rightarrow\ \Lambda_{c}^+ \bar{p} \pi^-$.
Points with errors and shaded histograms indicate the distributions
for the $B$ signal and the sideband regions, 
respectively. The curves indicate fits (see text for details).
}
\label{fig:fig2}
\end{figure}

Figure~\ref{fig:fig2} shows the $\Lambda_c^+\pi^{\pm}$
invariant mass distributions in the $B$
signal region, $|\Delta{E}|<0.030$~GeV and $M_{\rm{bc}}>5.270$~GeV/$c^2$.
Significant signals are observed for the $\Sigma_c(2455)$ and
$\Sigma_c(2520)$.
The shaded histograms are the distributions for events
in the sideband region
$0.040<|\Delta{E}|<0.100$~GeV,
normalized to the signal region $|\Delta{E}|<0.030$~GeV;
these account for continuum $\Sigma_c$ background.
The two curves indicate the results of separate fits to
the distributions for the $B$ signal and the sideband regions, 
with $\Sigma_c$ masses and
widths fixed to fit values for the signal MC events
generated with PDG values for masses and widths~\cite{pdg2000}.  
The background shapes are taken from a nonresonant signal MC.
To extract the $\Sigma_c$ yields, we performed
a simultaneous likelihood fit to the distributions for 
the $B$ signal and sideband regions.
We express the expected number $N_{\Sigma_c}$ of $B$ events as
$N_{\Sigma_c}=N_{Bb}-r \cdot N_{sb}$, where $N_{Bb}$ is the yield
in the $B$ signal region, $N_{sb}$ is the yield in the sideband region,
and $r=0.5$ is the normalization factor due to the ratio 
of their $\Delta{E}$ ranges, assuming a linear background shape.

Table~\ref{table:sigmac-minv} summarizes the observed signal yields and
branching fractions.
\begin{table*}[!htb]
\caption{
Efficiencies, yields, significances and branching fractions
for decay modes with $\Sigma_c^{++/0}$ resonances.
The errors are statistical, systematic, and a common error due to
the uncertainty in the value of 
${\cal{B}}(\Lambda_c^+\rightarrow{p K^- \pi^+})$.
 }
\begin{ruledtabular}
\begin{tabular}{lcccc}
Mode & Efficiency~(\%) & Yield & Significance & ${\cal{B}}~(\times10^{-4}$) \\ \hline
\hline
$\bar{B}^0\rightarrow\Sigma_c(2455)^{++}\bar{p}\pi^{-}$
& 4.93  & $18.6^{+4.9}_{-4.3}$ & 5.3 & $2.38^{+0.63}_{-0.55}\pm0.41\pm0.62$ \\
\hline
$\bar{B}^0\rightarrow\Sigma_c(2520)^{++}\bar{p}\pi^{-}$
& 6.38  & $16.5^{+5.8}_{-5.2}$ & 3.5 & $1.63^{+0.57}_{-0.51}\pm0.28\pm0.42$ \\
\hline
\hline
$\bar{B}^0\rightarrow\Sigma_c(2455)^{0}\bar{p}\pi^{+}$
& 4.80  & $6.4^{+3.2}_{-2.7}$ & 2.6 & $0.84^{+0.42}_{-0.35}\pm0.14\pm0.22$ \\
 & & \hspace{-5mm}$<~11.6$~(90\%~CL)\hspace{-5mm} & &
$<$~1.59~(90\%~CL) \\
\hline
$\bar{B}^0\rightarrow\Sigma_c(2520)^{0}\bar{p}\pi^{+}$
& 6.35  & $4.8^{+4.5}_{-4.0}$ & 1.2 & $0.48^{+0.45}_{-0.40}\pm0.08\pm0.12$ \\
 & & \hspace{-5mm}$<~11.7$~(90\%~CL)\hspace{-5mm} & &
$<$~1.21~(90\%~CL) \\
\hline
\hline
$B^-\rightarrow\Sigma_c(2455)^0\bar{p}$
& 6.00 & $4.3^{+2.5}_{-1.8}$ & 3.0 & $0.45^{+0.26}_{-0.19}\pm0.07\pm0.12$ \\
 & & \hspace{-5mm}$<~8.5$~(90\%~CL)\hspace{-5mm} & & $<~0.93$~(90\%~CL) \\
\hline
$B^-\rightarrow\Sigma_c(2520)^0\bar{p}$
& 7.47 & $1.7^{+1.8}_{-1.1}$ & 1.8 & $0.14^{+0.15}_{-0.09}\pm0.02\pm0.04$ \\
 & & \hspace{-5mm}$<~5.2$~(90\%~CL)\hspace{-5mm} & & $<~0.46$~(90\%~CL) \\
\end{tabular}
\end{ruledtabular}
\label{table:sigmac-minv}
\end{table*}
We observe the $\bar{B}^0\rightarrow\Sigma_c(2455)^{++}\bar{p}\pi^-$  
decay for the first time with a statistical
significance of 5.3$\,\sigma$. 
We also see 3.5$\,\sigma$ evidence 
for $\bar{B}^0\rightarrow\Sigma_c(2520)^{++}\bar{p}\pi^-$,  
2.6$\,\sigma$ evidence for $\bar{B}^0\rightarrow\Sigma_c(2455)^0\bar{p}\pi^+$,  
and less evidence for $\bar{B}^0\rightarrow\Sigma_c(2520)^0\bar{p}\pi^+$.
We see 3.0$\,\sigma$ evidence for the two-body decay 
$B^-\rightarrow\Sigma_c(2455)^0\bar{p}$, and less evidence 
for $B^-\rightarrow\Sigma_{c}(2520)^0\bar{p}$. 
For those modes with a significance of three sigmas or less,
we set upper limits on their branching fractions.

Our results provide stringent constraints upon
theoretical predictions~\cite{jarfi,chernyak,ball,cheng}. 
The predictions for 
$\bar{B}^0\rightarrow\Lambda_c^+\bar{p}$ in \cite{jarfi,chernyak,ball}
were already much larger than the CLEO experimental upper limit~\cite{cleo-lamc}; 
here we set an even more restrictive upper limit.
A recent study based on a bag model~\cite{cheng} gives predictions of branching 
fractions of $\le(0.1\sim0.3)\times10^{-4}$ for $\bar{B}^0\rightarrow\Lambda_c^+\bar{p}$ 
and $(4.3\sim15.1)\times10^{-4}$ for $B^-\rightarrow\Lambda_c^+\bar{p}\pi^-$.
Our upper limit for $\bar{B}^0\rightarrow\Lambda_c^+\bar{p}$
does not contradict this model,
while our measured result for $B^-\rightarrow\Lambda_c^+\bar{p}\pi^-$ 
is much smaller than its predicted value.

In summary, 
we have observed the exclusive three-body decay 
$\bar{B}^0\rightarrow\Sigma_{c}(2455)^{++}\bar{p}\pi^-$ for the first time
and observed evidence for the exclusive two-body decay
$B^-\rightarrow\Sigma_c(2455)^0\bar{p}$.
We make improved measurements of the branching fractions
for $\bar{B}^0\rightarrow\Lambda_c^+\bar{p}\pi^+\pi^-$ and
$B^-\rightarrow\Lambda_c^+\bar{p}\pi^-$, and 
also set a more restrictive upper limit 
on $\bar{B}^0\rightarrow\Lambda_c^+\bar{p}$.

\begin{acknowledgments}
We wish to thank the KEKB accelerator group for the excellent
operation of the KEKB accelerator.
We acknowledge support from the Ministry of Education,
Culture, Sports, Science, and Technology of Japan
and the Japan Society for the Promotion of Science;
the Australian Research Council
and the Australian Department of Industry, Science and Resources;
the National Science Foundation of China under Contract No. 10175071;
the Department of Science and Technology of India;
the BK21 program of the Ministry of Education of Korea
and the CHEP SRC program of the Korea Science and Engineering Foundation;
the Polish State Committee for Scientific Research
under contract No.~2P03B 17017;
the Ministry of Science and Technology of the Russian Federation;
the Ministry of Education, Science and Sport of Slovenia;
the National Science Council and the Ministry of Education of Taiwan;
and the U.S. Department of Energy.
\end{acknowledgments}

\end{document}